\documentclass[%
 aip,
 jmp,%
 amsmath,amssymb,
preprint,%
]{revtex4-1}
\usepackage{graphicx}
\usepackage{dcolumn}
\usepackage{bm}
\usepackage{hyperref}

\usepackage{color}
\input xy
\xyoption{all}

\newcommand{\be}{\begin{equation}}
\newcommand{\ee}{\end{equation}}
\newcommand{\bea}{\begin{eqnarray}}
\newcommand{\eea}{\end{eqnarray}}
\newcommand{\bd}{\begin{displaymath}}
\newcommand{\ed}{\end{displaymath}}

\newcommand{\qi}{ q^{-1}}

\newcommand{\ad }{a^{\dagger}}

\newcommand{\h }{ \hbar}

\newcommand{\p}{ \partial_x }

\newcommand{\e }{ {\cal E}}

\newcommand{\lb }{ \left( }
\newcommand{\rb }{ \right ) }

\begin{document}

\title
{Generalized 
$q-$deformed Tamm-Dancoff oscillator algebra 
and associated coherent states }
\author{ Won Sang Chung${}^1$ }
\email{mimip4444@hanmail.net}

\author{Mahouton Norbert Hounkonnou${}^2$}
\email{norbert.hounkonnou@cipma.uac.bj}

\author{Sama  Arjika${}^2$}
\email{rjksama2008@gmail.com}

\affiliation{
${}^1$Department of Physics and Research Institute of Natural Science, \\
College of Natural Science, Gyeongsang National University, Jinju 660-701, Korea,
\\
${}^2$International Chair in Mathematical Physics and Applications\\
(ICMPA--UNESCO Chair), 072 B.P. 50  Cotonou, Republic of Benin\\
University of Abomey-Calavi
}

\date{\today}

\begin{abstract}
In this paper, we  propose a   full characterization of a generalized $q-$deformed Tamm-Dancoff
oscillator algebra and  investigate its main mathematical and physical properties.
Specifically, we  study its various  representations
and find the 
condition satisfied by  the deformed $q-$number to define the algebra structure function. Particular Fock spaces involving   finite and infinite dimensions are examined.  A
 deformed calculus is performed as well as  a coordinate realization for this 
algebra. A relevant example
is exhibited. Associated
 coherent states are constructed.
Finally,   some thermodynamics aspects are computed and discussed.
\end{abstract}

\maketitle

\section{Introduction}
Quantum algebras and quantum groups 
play a leading role
in physics and mathematics. 
Quantum 
groups or $q-$deformed Lie algebras imply some specific deformations
 of classical Lie algebras. From the mathematical point of view, it is a non-commutative
 associative Hopf algebra. The structure and representation theory of quantum 
groups have been developed extensively by Jimbo \cite{M. Jimbo} and Drinfeld \cite{V. Drinfeld} (and references therein).

The $q-$deformation of the oscillator algebra was first  accomplished by Arik 
and Coon \cite{M. Arik} and lately accomplished by 
Macfarlane \cite{A. Macfarlane} and Biedenharn \cite{L. Biedenharn} by 
using the $q-$calculus which was originally introduced by Jackson 
in the early 20th century \cite{F. Jackson}. 
As matter of other relevant works citation, let us also mention the $q-$oscillator
 algebras investigated by Kuryshkin~\cite{Kuryshkin80}, 
Jannussis and  collaborators~\cite{Jannussis&al83b}, Hounkonnou et {\it al}  \cite{Hounkonnou&Ngompe07b,Baloitcha} and references therein.
In the study of the basic
hypergeometric functions, Jackson invented the Jackson derivative 
and integral, which is now called $q-$derivative and $q-$integral. Jackson's pioneering 
research enabled theoretical physicists and mathematicians to study 
new physics and mathematics related to the $q-$calculus. Much was 
accomplished in this direction and work is under way to
 find the meaning of the deformed theory.

Historically, the $q-$deformed Tamm-Dancoff
oscillator algebra was first introduced in \cite{K. Odaka}, and some of
its Hopf algebraic aspects were also discussed in \cite{S. Chaturvedi}.
We designate here this model under the name of  the TD-oscillator model. It should be pointed out
that some of the quantum statistical properties of this model, with the range
$q<1,$ have been also considered in \cite{A. M. Gavrilik1,A. M. Gavrilik2} 
in the investigations 
on the two-parameter-deformed oscillators.

The TD-oscillator  model is defined by the following commutation relations
\be
\label{algebra:1}
aa^{\dagger} - q  a^{\dagger}a=q^{N},~~~
[ N, a^{\dagger} ] = a^{\dagger}, ~~ [N, a]= -a,
\ee
where $a$, $a^\dagger$ and $N$ are the  annihilation, creation and number operators, respectively.
The algebra (\ref{algebra:1}) was shown to have a Hopf algebra 
structure
\cite{S. Chaturvedi}.
See also \cite{W. S. Chung}, where  is shown  
the Hopf algebra structure of a generalized Heisenberg-Weyl algebra.

In this paper, we consider a generalization of the $q-$deformed 
Tamm-Dancoff oscillator algebra and investigate its main mathematical and physical properties.

The paper is organized as follows.  In Section II, we study the representation 
for the generalized $q-$deformed TD oscillator algebra and find the 
condition satisfied by the deformed number.
In Section III, we find the 
deformed derivative and deformed integral and obtain a 
coordinate realization for this algebra. An interesting  example 
is discussed. 
Associated
coherent states are constructed
in the Section IV.  Section V is devoted to
the generalized $q-$deformed TD oscillator algebra in
$d-$dimensional Fock space. Its representation as well as the 
deformed derivative and deformed integral are given. 
Finally, some thermodynamics aspects are discussed in Section VI.  
\section{Generalization of the q-deformed Tamm-Dancoff oscillator algebra}
Let us consider the following algebra
\be
\label{deformed:al}
a\ad - \ad a = \{ N+1 \} - \{ N \} , ~~~
[ N, a^{\dagger} ] = a^{\dagger}, ~~ [N, a]= -a,
\ee
where the new $q-$deformed  number is defined as
\be
\label{number:d}
\{ N \} = N ( \mu q^{ \alpha N + \beta } + \eta 
q^{ \gamma N + \delta }), ~~~ \alpha >0,~ \alpha \ne \gamma, ~ q > 0.
\ee
Meljanac et {\it al}
 \cite{Meljanac} introduced the 
generalized $q-$deformed single-mode oscillator algebra 
through the identity operator $\mathbf{1}$,
a self-adjoint number operator $N$,  a lowering operator $a$ 
and  an  operator $\bar{a}$ which is not necessarily conjugate to $a$
 satisfying
\bea
 && [N,\;a]= -a, \quad 
[N,\;\bar{a}]= \bar{a}, \label{uq2}\\
&& a\bar{a}-F(N)\bar{a}a=G(N),  \label{uq3}
\eea
where $F$ and $G$ are arbitrary complex analytic functions. The same algebra was
 investigated by  Bukweli and Hounkonnou \cite{Bukweli}. These authors show that   
 from the relation (\ref{uq2}), one can get
\be 
[N,\;a\bar{a}]= 0=[N,\;\bar{a}a]
\ee
implying the existence of a complex analytic function $\varphi$ such that
\be 
 \bar{a}a=\varphi(N)\quad\mbox{ and }\quad a\bar{a}=\varphi(N+1).\label{uq4}
\ee
Therefore, (\ref{uq3}) can be rewritten as follows:
\be 
\label{wsc:uq5}
\varphi(N+1)-F(N)\varphi(N)= G(N),
\ee
where the structure function $\varphi(n)$ is as follows \cite{Bukweli}
\be
\varphi(n)= [F(n-1)]!\sum_{k=0}^{n-1}\frac{G(k)}{[F(k)]!},\quad n\geq 1,
\ee
and
\be 
 [F(k)]!=\left\{\begin{array}{lcl}F(k)F(k-1)\cdots F(1)&\mbox{ if }& k\geq 1\\1&\mbox{ if }& k=0. \end{array}\right.
\ee
Let us denote now $a^\dagger$ the Hermitian conjugate of the operator $a$. Then,
\be 
[N,\;a^\dagger]= a^\dagger\quad\mbox{ and }\quad \bar{a}=c(N)a^\dagger,
\ee
where $c(N)$ is   given by $c(N)= e^{i\arg{\varphi(N)}}$.
The algebra (\ref{deformed:al}) is the  particular case of (\ref{wsc:uq5})
with $\varphi(N)=\{ N \},\;F(N)=q$ and $G(N)= \mu q^{ \alpha N + \beta } + \eta q^{ \gamma N + \delta }.$\\
If we choose $ \mu =1, \eta =0 , \alpha =1, \beta =-1$, the algebra
 (\ref{deformed:al}) becomes TD-algbera. The algebra (\ref{deformed:al}) is 
called a generalized $q-$deformed Tamm-Dancoff oscillator algebra.
When $q$ goes to unity, we have $ \{ N \} = ( \mu + \eta )N$. For correspondence, we demand
\be
 \mu + \eta  = 1.
 \ee
 Then the relation (\ref{number:d}) becomes
 \be
\{ N \}_\mu = N ( \mu q^{ \alpha N + \beta } + ( 1 - \mu ) q^{ \gamma N + \delta }).
\ee
 From now  we  restrict our concern to the case when $ q $ is real. We are interested in the Fock 
representation of the algebra (\ref{deformed:al}); this is an irreducible representation constructed
 on a Hilbert space with the orthonormal basis of vectors $|n\rangle , ~ n =0, 1, 2, \cdots $.
The action of $N$ is standard in the sense that
\be
N|n\rangle = n |n\rangle,~~~n=0,1,2,\cdots ,
\ee
while the action of the remaining operators is given by
\bea\label{norm}
a|n\rangle &=& \sqrt{\{ n \}_\mu }\, |n-1\rangle\\
a^{\dagger} |n\rangle  &=&\sqrt{\{ n+1 \}_\mu}\, |n+1\rangle.
\eea
The latter relations 
require $ \{ n \}_\mu \ge 0 $, implying
\be
\label{ineq:li}
q^{ \alpha n + \beta } \ge \left( 1 - \frac {1}{\mu } \right) q^{ \gamma n + \delta }.
\ee
When $\mu  \le 1$, the inequality holds for all $n$. However, it is not the case 
when $ \mu >1 $. In this case, the solution of the inequality (\ref{ineq:li}) depends 
on the value of $q$ and $\alpha - \gamma$. Thus, we have the following four cases.
\begin{itemize}
\item {\bf Type I :} $ q > 1 , \alpha > \gamma $
\be
n \ge \frac{1}{ \alpha - \gamma } \left[  \delta - \beta + \log_q \lb 1 - \frac{1}{\mu} \rb \right]
\ee
\item {\bf Type  II :} $ q > 1 , \alpha < \gamma $
\be
n \le \frac{1}{ \alpha - \gamma } \left[  \delta - \beta + \log_q \lb 1 - \frac{1}{\mu} \rb \right]
\ee
\item  {\bf Type III :} $ 0< q < 1 , \alpha > \gamma $
\be
n \le \frac{1}{ \alpha - \gamma } \left[  \delta - \beta + \log_q \lb  1 - \frac{1}{\mu} \rb \right]
\ee
\item {\bf Type IV :} $ 0< q < 1 , \alpha < \gamma $
\be
n \ge \frac{1}{ \alpha - \gamma } \left[  \delta - \beta + \log_q \lb 1 - \frac{1}{\mu} \rb \right].
\ee
\end{itemize}
\section{ The $q-$deformed Tamm-Dancoff oscillator
 algebra with an infinite dimensional Fock space}

In this section we suggest an interesting example for the algebra (\ref{deformed:al}) with 
infinite dimensional Fock space. We restrict our concern to the case when $ 0<q<1 $. Let us take the following values:
\bd
\alpha =-1,\quad  \beta =  1,\quad  \gamma =1,\quad  \delta =-1.
\ed
With  this choice, we have
\be
\label{eq13}
a\ad - \ad a = \{ N+1 \}_\mu - \{ N \}_\mu, ~~~
[ N, a^{\dagger} ] = a^{\dagger}, ~~ [N, a]= -a,
\ee
where
\bd
\ad a = \{ N \}_{\mu} = N \lb \mu\,  q^{-N+1} +(1 -\mu )  q^{N-1} \rb.
\ed
This choice gives us an infinite dimensional representation. Therefore, we have
\bea
N|n\rangle &=& n |n\rangle, ~~~~ n=0, 1, 2, \cdots,\\
a|n\rangle &=& \sqrt{  n \lb \mu q^{ -n +1} + ( 1 - \mu ) q^{n-1}  \rb  }\,|n-1\rangle,\\
\ad |n\rangle& =& \sqrt{ (n+1) \lb \mu q^{ -n } + ( 1 - \mu ) q^{n}   \rb  }\,|n+1\rangle.
\eea
In order to have a functional realization of this representation, we consider 
the space $\cal{P}$ of all monomials in variable $x$, and introduce its basis of monomials
\be
|n \rangle: = \frac{x^n}{ \sqrt{ \{n\}_{\mu} ! }},
\ee
where
\be
\{n\}_{\mu} ! = \prod_{k=1}^n \{ k \}_{\mu} , ~~~~ \{ 0 \}_{\mu} ! =1.
\ee
Then, the functional realization of the algebra (\ref{eq13}) is given by
\be
a : = D_x , ~~~~ \ad: = x , ~~~ N : = x \p ,
\ee
where the new deformed derivative is given by
\be
\label{newderivati}
D_x =  \lb \mu q^{ -x \p   } + ( 1 -\mu)  q^{ x \p } \rb \p
=  \lb \mu T_q^{-1} + ( 1 - \mu ) T_q \rb \p,
\ee
and
\be
 T_q f(x) = f(qx).
 \ee
 The Leibniz rule of the deformed derivative is then given by
\bea
D_x ( f(x) g(x) ) &=& ( D_x f(x) ) (T_q^{-1} g(x) )+( T_q f(x)) ( D_x g(x) )
- \mu ( {\cal T }f(x) ) ( T_q^{-1} \p g(x) )\cr
&+& ( 1 - \mu ) ( T_q \p f (x)) ( {\cal T }g(x) ),
\eea
where
\be
{\cal T }f(x) = ( T_q - T_q^{-1} ) f(x) = f( qx ) - f( q^{-1} x ).
\ee
Let  $Q_\mu $ and
$P_\mu $ be the deformed position and momentum operators defined as follows:
\be
Q_\mu:= \left(1/{2 \, m}\,\omega\right)^{1/2}(a^\dag+a)\quad 
\mbox{ and } \quad P_\mu:=i \left(m \, \omega/{2}\right)^{1/2}(a^\dag-a).
\ee
 The operators $(a^\dag+a)$
and
$i(a^\dag-a)$
are not essentially
 self-adjoint, but
have a one-parameter family of
self-adjoint extensions  for $0< q<1$.

Indeed, 
the matrix elements of  the operator
$  a^\dag+a$ on  the basis vector $|n\rangle$ of the space $\cal{P}$ 
 are given by
\be
\langle m|  a^\dag+a|n\rangle
=
b_{n,\mu}\,\delta_{m,n+1}+
b_{n-1,\mu}\,\delta_{m,n-1}, \quad n,\;m=0,
\;1,\;2,\;\cdots,
\ee 
while the matrix elements of the operator
$\;i (a^\dag-a)$  are given by
\be 
\langle m|i (a^\dag-a)|n\rangle
=
i b_{n,\mu}\,\delta_{m,n+1}-i b_{n-1,\mu}\,
\delta_{m,n-1},\quad n,\;m=0,\;1,\;2,\;\cdots,
\ee 
where  $b_{n,\mu}= \{n+1\}_\mu$. Besides,
the   operators $(a^\dag+a)$ and  $i(a^\dag-a)$
   can be represented by the two following symmetric Jacobi matrices, respectively,
\bea\label{sama:jacobir}
M_{Q_\mu}= \left(\begin{array}{ccccccc}0&
b_{0,\mu}&0&0&0&0&\cdots\\b_{0,\mu}
&0&b_{1,\mu}&0&0&0&\cdots\\0
&b_{1,\mu}&0&b_{2,\mu}&0&0&\cdots\\
0
&0&b_{2,\mu}&0&b_{3,\mu}&0&\cdots\\\vdots&\ddots&
\ddots&\ddots&\ddots&\ddots&\ddots
       \end{array}\right)
\eea
and
\bea\label{sama:jacobic}
M_{P_\mu}= \left(\begin{array}{ccccccc}0&-ib_{0,\mu}
&0&0&0&0&\cdots\\ib_{0,\mu}
&0&-ib_{1,\mu}&0&0&0&\cdots\\0&ib_{1,\mu}&0
&-ib_{2,\mu}&0&0&\cdots
\\
0
&0&ib_{2,\mu}&0&-ib_{3,\mu}&0&\cdots\\\vdots&
\ddots&\ddots&\ddots&\ddots&
\ddots&\ddots
       \end{array}\right).
\eea
For $0 <q <1,$ we have
\be
 \lim_{n\to\infty}b_{n,\mu}=\lim_{n\to\infty}
\left((n+1)(\mu q^{-n}+(1-\mu)q^n)\right)^{1/2}=\infty.
\ee
Considering the series
$\sum_{n=0}^\infty 1/b_{n,\mu}$, we obtain
\be
\overline{\lim_{n\to\infty}}\left(\frac{1/b_{n+1,\mu}}{1/b_{n,\mu}}\right)=
\overline{\lim_{n\to\infty}}\left(\frac{(n+1)(\mu q^{-n}+(1-\mu)q^n)}{(n+2)(\mu q^{-n-1}+(1-\mu)q^{n+1})}
\right)^{1/2}= q^{1/2}<1.
\ee
This ratio test  leads to the conclusion  that the series
$\sum_{n=0}^\infty 1/b_{n,\mu}$ converges.
Moreover, 
$1-2q^{2}+q^4=(1-q^{2})^2>0$
$
\Longrightarrow    2 \leq q^{-2}+q^2$ and 
  $2\mu(1-\mu) \leq \mu(1-\mu) (q^{-2}+q^2)<2\mu(1-\mu)+\mu^2q^{-2}+(1-\mu)^2q^{2}.$
 Hence,
\bea
0\leq b_{n-1,\mu}\,b_{n+1,\mu}&=&(n^2+2n)(\mu^2 q^{-2n}+\mu(1-\mu)(q^2+q^{-2})+(1-\mu)^2q^{2n})\cr
&\leq &(n^2+2n)(\mu^2 q^{-2n}+\mu(1-\mu) +(1-\mu)^2q^{2n})\cr
&+& (\mu^2 q^{-2n}+\mu(1-\mu) +(1-\mu)^2q^{2n})= b_{n,\mu}^2.
\eea
Therefore, the Jacobi matrices  in (\ref{sama:jacobir}) and
(\ref{sama:jacobic}) are not self-adjoint (Theorem 1.5., Chapter VII in Ref. \cite{Berezanskii})
but have each a one-parameter family of self-adjoint extensions.

Let 
\bea
H_\mu:
&=&\frac{1}{2m}(P_\mu)^2+\frac{1}{2}m\omega^2(Q_\mu)^2\cr
&&\cr
&=&\frac{ \omega}{2}(a^\dagger a+aa^\dagger)
\eea
 be the  deformed Hamiltonian associated
 to the algebra (\ref{eq13}). The following statement holds:
\begin{itemize}
\item The vectors $|n\rangle$
are eigen-vectors of   $H_\mu$ with respect to the eigenvalues
\be
\label{mecp1}
 E_{\mu}(n)=  \frac{  \omega}{2}\Big(\{n\}_\mu+ \{n+1\}_\mu\Big),
\ee
\item The mean values of $\;Q_\mu$ and $P_\mu$ in the states $|n\rangle$ are zero while
their variances  are given by
\be 
 (\Delta  Q_\mu)_{n}^2= \frac{1}{2m\omega}\Big(\{n\}_\mu+ \{n+1\}_\mu\Big)
\ee
and
\be
  (\Delta P_\mu)_{n}^2= \frac{  m \omega}{2}\Big(\{n\}_\mu+ \{n+1\}_\mu\Big),
\ee 
respectively, where $(\Delta X)_n^2=\langle X^2\rangle_n-\langle X\rangle_n^2$
with $\langle X\rangle_n=\langle n|X|n\rangle$.
\item The position-momentum uncertainty relation is given by
\be
 (\Delta Q_\mu)_{n}(\Delta P_\mu)_{n}=\omega^{-1}E_{\mu}(n),
\ee
which is reduced, for the vacuum state,  to the expression
\be
 (\Delta Q_\mu)_{0}(\Delta P_\mu)_{0}= \frac{ 1}{2}.
\ee
\end{itemize}
Let us turn back to the  derivative (\ref{newderivati}) and defined the
   deformed integral  as follows:
\bea
\int Dx f(x) &:=& \int dx  \,( \mu T_q^{-1} + ( 1 - \mu ) T_q )^{-1} f(x)\cr
&=& \mu^{-1} \sum_{n=0}^{\infty} \lb 1 - \frac{1}{\mu} \rb^n \int dx f( q^{2n +1 } x ).
\eea
Applying the deformed derivative and the deformed integral to  $x^n $ yields
\bd
D_x\, x^n = \{ n \}_{\mu} x^{n-1} ~~~ \mbox{ and }~~~ \int Dx\, x^n = \frac{x^{n+1}}{\{ n+1 \}_{\mu} }.
\ed
For the deformed derivative and the deformed integral, we have the following formulae
\be
\int D x\, \frac{1}{x}  = \frac{1}{ \mu q + ( 1 - \mu ) q^{-1} }  \ln x  ~~~ \mbox{ and }~~~ D_x (  \ln x ) = \frac{ \mu q + ( 1 - \mu ) q^{-1}}{x},
\ee
where $ \int D x\, \frac{1}{x}  $ exists for $ q > \sqrt { 1 - \frac{1}{\mu}} $.

The deformed exponential function $\e_{\mu} (x) $ is defined as
\be
\label{new:expo}
\e_{\mu} (x) := \sum_{n=0}^{\infty} \frac{1}{\{ n \}_{\mu}!} x^n,
\ee
  satisfying
\be
\label{new:dif}
D_x \,\e_{\mu} ( \omega\, x ) = \omega \,\e_{\mu} ( \omega\,x )
\ee
for an arbitrary constant $\omega$. From the relation
\be
\int_0^{\infty} Dx\, \e_{\mu} ( -\omega\, x ) = \frac{1}{\omega},
\ee
we can obtain the following formula
\be
\label{int:new}
\int_0^{\infty} Dx \,\e_{\mu} ( -\omega\, x )\, x^n = \frac{(-1)^n }{\omega^{n+1}}\prod_{k=1}^n \{ - k \}_{\mu}
\ee
Inserting $\omega=1$ into the relation (\ref{int:new}) yields
\be
\label{int:newq}
\int_0^{\infty} Dx\, \e_{\mu} ( - x ) \,x^n = (-1)^n \prod_{k=1}^n \{ - k \}_{\mu}.
\ee
Replacing $\mu$ with $ 1- \mu $ in   the relation  (\ref{int:newq}), we have
\be
\label{moment:pr}
\int_0^{\infty} Dx\, \e_{1-\mu} ( - x ) x^n =\frac{ 2\, \{ n+2\}_{\mu} !}{ \{ 2 \}_{\mu} ( n+1) ( n+2 ) },
\ee
where we use
\be
\{- k \}_{\mu} = - \frac{ k}{k+2} \{ k+2 \}_{1- \mu }.
\ee
\section{ Coherent states}
In this section, we construct the coherent states of the generalized 
TD-oscillator algebra (\ref{eq13}). The coherent states $|z\rangle$ are defined as   
the eigenstates of the annihilation operator in the form
\be
\label{ani:h}
a |z\rangle := z |z\rangle.
\ee
They  can be   represented by using the eigenvector of the number operator as follows :
\be
\label{ani:he}
|z \rangle = \sum_{n=0}^{\infty} c_{n} (z) |n\rangle.
\ee
Inserting the relation (\ref{ani:he}) into (\ref{ani:h}), we have
\be
\label{ani:hew}
\sum_{n=1}^{\infty } c_{n} (z) \sqrt{ \{ n \}_{\mu} } \,|n-1\rangle = \sum_{n=0}^{\infty } z\,c_{n} (z) \,|n\rangle.
\ee
From   (\ref{ani:hew}), we get the following recurrence relation
\be
\label{ani:heww}
c_{ n+1} (z)= \frac{z}{\sqrt{ \{ n +1\}_{\mu} } } c_n(z), ~~~   n=0, 1, 2, \cdots
\ee
giving
\be
c_n(z)= \frac { z^n} { \{ n \}_{\mu}! }  c_0(z).
\ee
From $\langle z |z\rangle =1 $, we have
\be
c_{0}^{-2}(z) = \e_{\mu} ( |z|^2 )
\ee
and the coherent states (\ref{ani:he}) become
\be
\label{ani:coh}
|z \rangle = \e_{\mu}^{-1/2} (|z|^2)\sum_{n=0}^{\infty} \frac{z^n}{\sqrt{\{ n \}_{\mu}! }} |n\rangle.
\ee
They are continuous  in their label $z$.
Indeed,
\be
\left\| \vert z\rangle - \vert
z'\rangle\right\|^2 = 2\left(1 -
Re(\langle z\vert z'\rangle)\right),
\ee
where
\be
\langle z\vert z'\rangle=\left(\e_{\mu}  (|z'|^2)\e_{\mu} (|z|^2) \right)^{-1/2}\e_{\mu} (z\bar{z'}).
\ee
So,
\be
\left\|\vert
z\rangle - \vert z'\rangle
\right\|^2 \to  0  \mbox{ as } |z-z'|\to 0,
\mbox{ since } \langle z\vert z'\rangle\to 1\mbox{ as }\vert z - z'\vert \to   0.
\ee
Besides, 
 their  overcompleteness relation can be established as follows:
\be
\label{sahonkar}
\frac{1}{ \pi } \int \int |z\rangle \mu( |z|^2 ) \langle z | |z| D|z| d \theta = I,
\ee
where $ \mu ( |z|^2 ) $ is a weight function. Inserting (\ref{ani:coh}) into (\ref{sahonkar}), we obtain
\be
\sum_{n=0}^{\infty} \frac{ 1}{ \{ n \}_\mu !   }
|n\rangle \langle n | \int_0^{\infty} \frac{\mu (x)}{ \e_{\mu}(x)}  x^n Dx = I ,
\ee
where $ x = |z|^2,$
imposing to find a function $f$ such that
\be
\label{saa:chungw}
\int_0^{\infty} f(x)  x^n D x =  \{ n \}_\mu! .
\ee
Therefore, not  all deformed algebras lead to 
coherent states since the moment problem (\ref{saa:chungw}) 
does  not always have solution \cite{Akhiezer,Tarmakin}.
So, the question arises  is then how to determine the   function $f(x)$ in (\ref{saa:chungw})? 
Using the formula (\ref{moment:pr}), we can write
\be
\label{saa:chung}
f(x) = \e_{1-\mu} ( -x ) g(x) = \e_{1-\mu} (-x) \sum_{k=0}^{\infty} g_k x^k.
\ee
Inserting   (\ref{saa:chung}) into   (\ref{saa:chungw}) yields
\be
\sum_{k=0}^{\infty} g_k \frac{ ( n+k)!}{n!} \prod_{j=n+1}^{n+k+2} \phi(j) = \phi (2),
\ee
where
\be
\phi(j) = \frac{ \{ j \}_{\mu} }{ j }.
\ee
Solving the above equation, we have
\bd
g_k = \frac{ (-1)^k }{k! \phi ( k+2 ) ! } \left[ \phi (2) - \sum_{i=0}^{k-1} g_i 
\frac{ (-1)^i k! }{ ( k-i )! } \prod_{j= k - i +1 }^{ k+2 } \phi (j) \right], ~~k \ge 1 
\ed
with 
\be
g_0 =1 , ~~\phi(i)! = \prod_{j=1}^i \phi(j).
\ee
The first few $ g_k $'s are as follows:
\bea
g_0 &=&1\\
g_1 &=& \frac{ \phi (2) ( \phi (3) - 1 ) }{\phi(3)! }\\
g_2& =& \frac{ \phi (2) - \phi(3) \phi(4) + 2 \phi (2) ( \phi (3) - 1 ) \phi(4) }{2!\,\phi(4)! }.
\eea
On the other way, using the formula (\ref{moment:pr}), we have
\be
 \phi (2)\int_0^{\infty} Dx\, \e_{1-\mu} ( - x ) (\p)^2\,x^{n+2} =  \{ n+2\}_{\mu}!.
\ee
By replacing $n+2$ by $n$, the latter equation is equivalent to
\be
\int_0^{\infty} Dx\left( \phi (2)(\p)^2\e_{1-\mu} ( - x )\right) \,x^{n} =  \{ n\}_{\mu}!,
\ee
with $\e_{1-\mu} ( - x ) \p x^{n}\,\Big|_0^\infty=0=
\p \e_{1-\mu} ( - x )  \,x^{n}\,\Big|_0^\infty.$
Therefore, the function $f(x)$ has the form
\be
f(x)= \phi (2)  \left(\p\right)^2\e_{1-\mu} ( - x )
\ee
from which the relation
\be
\mu(x)= \phi (2)\, \e_{\mu} ( x )  \left(\p\right)^2\e_{1-\mu} ( - x )
\ee
follows.
\section{ $d$-dimensional }
In this section we suggest an interesting example for the 
algebra (\ref{number:d}). Now we restrict our concern to the 
case that $ 0<q<1 $. Let us take the following values:
\bd
\alpha =1, \,\beta =  -1,\, \gamma =-1, \,\delta =-1.
\ed
We also set
\be
\log_q \lb  1 - \frac{1}{\mu} \rb = 2 d
\ee
where $ d \in Z_+ $. In this choice, we have
\be
\label{sa:asa}
\ad a = \{ N \}_d = N \lb \frac{1}{1-q^{2d} } q^{N-1} - \frac{q^{2d} }{1-q^{2d} } q^{-N-1} \rb.
\ee
We can easily find that  (\ref{sa:asa}) reduces to the Tamm-Dancoff 
case when $ d$ goes to infinity. This choice gives us the 
$d$-dimensional representation of algebra (\ref{deformed:al}):
\bea
N|n\rangle& =& n |n\rangle, ~~~~ n=0, 1, 2, \cdots , d-1\\
a|n\rangle &=& \sqrt{ q^{-1} n \lb \frac{ q^n - q^{2d-n}} { 1-q^{2d} } \rb  }\;|n-1\rangle,\\
\ad |n\rangle &=& \sqrt{ q^{-1} (n+1) \lb \frac{ q^{n+1} - q^{2d-n-1}} { 1-q^{2d} } \rb  }\;|n+1\rangle.
\eea
With these considerations, the functional realization of the algebra (\ref{eq13}) is also given by
\be
a := D_x^d , ~~~~ \ad  := x , ~~~ N  := x \p ,
\ee
where the   deformed derivative $D_x^d$ is given by
\be
\label{ewderivati}
D_x^d =  \frac{   q^{ x \p   } -  q^{2d-2-x \p }}{1-q^{2d}} \p
=   \frac{    T_q  -q^{2d-2}\, T_q^{-1}}{1-q^{2d}} \p.
\ee
 The Leibniz rule of the deformed derivative is then given by
\bea
D_x^d ( f(x) g(x) ) &=& ( D_x^d f(x) ) (T_q  g (x))+( T_q^{-1} f(x)) ( D_x^d g(x) )
+ ( {\cal T }_df(x) ) ( T_q  \p g(x) )\cr
&+&  q^{2d-2} ( T_q^{-1} \p f(x) ) ( {\cal T }_d\,g(x) ),
\eea
where
\be
{\cal T }_d\,f(x) = \frac{ T_q - T_q^{-1} }{1-q^{2d}} f(x) = \frac{f( qx ) - f( q^{-1} x )}{1-q^{2d}}.
\ee
Let us turn back to the  derivative (\ref{ewderivati}) and defined the
   deformed integral  as follows:
\bea
\int  D^dx f(x) &=&(1-q^{2d})  \int dx  (   T_q   -q^{2d-2} T_q^{-1} )^{-1} f(x)\cr
&= &q^{2-2d}(q^{2d}-1) \sum_{n=0}^{\infty} q^{n(2-2d)} \int dx f( q^{2n +1 } x ).
\eea
Applying the deformed derivative and the deformed integral to  $x^n $ yields
\bd
D_x^d\, x^n = \{ n \}_d\, x^{n-1}~~~\mbox{ and } ~~~\int D^dx\, x^n = \frac{x^{n+1}}{\{ n+1 \}_d}.
\ed
As a particular case, we have
\be
\int  D^dx \,\frac{1}{x}  = q \ln x , ~~~ D_x^d (  \ln x ) = \frac{   q^{-1}}{x}.
\ee
The deformed exponential function $E_{d} (x) $  can be defined as:
\be
\label{new:expo}
E_d (x) := \sum_{n=0}^{\infty} \frac{1}{\{ n \}_d!} x^n,
\ee
where
\be
\{n\}_d ! = \prod_{k=1}^n \{ k \}_d , ~~~~ \{ 0 \}_d ! =1
\ee
with the property
\be
\label{ne:dif}
D_x^d \,E_d ( \omega\, x ) = \omega \,E_d ( \omega \,x )
\ee
for an arbitrary constant $\omega$. From the relation
\be
\int_0^{\infty} D^dx\, E_d ( -\omega \,x ) = \frac{1}{\omega},
\ee
we can derive the following formula
\be
\label{in:new}
\int_0^{\infty}  D^dx\,E_{d} ( -\omega \,x ) x^n = \frac{(-1)^n }{\omega^{n+1}}\prod_{k=1}^n \{ - k \}_d.
\ee
Setting  $\omega=1$ into the relation (\ref{in:new}) yields
\be
\label{in:newq}
\int_0^{\infty}  D^dx\, E_{d} ( - x ) x^n =    \{n \}_{-d}!
\ee
where we use
\be
\{- k \}_{d} = -   \{ k  \}_{-d }.
\ee
\section{Deformed black body radiation}
The thermodynamics properties are shown to be 
determined by the partition function $Z$ defined by
\be
Z= Tr (e^{-\beta H} )  = \sum_{n=0}^{\infty} \langle n| e^{-\beta H} |n\rangle,
\ee
where $ \beta = 1/kT $.
In the generalized TD (GTD)-oscillator algebra, we assume the Hamiltonian as
\be
\label{ognon}
H:= w N.
\ee
Now we can compute the partition function for the GTD-oscillator as follows:
\be
Z= \sum_{n=0}^{\infty} \langle n| e^{-\beta H } |n\rangle = \frac {1 }{1- e^{-\beta w }}.
\ee
For any operator $\hat{O},$  the ensemble average is then defined by
\be
\langle \hat{O} \rangle := \frac {1}{Z} Tr ( e^{- \beta H } \hat{O} ).
\ee
While the thermodynamics for a system with Hamiltonian (\ref{ognon}) is
 independent of the deformation, Green functions like 
$\langle \ad a \rangle$ will depend on the deformation. For the symmetric Tamm-Dancoff (STD)-oscillator, we have
\be
\label{swh}
\langle \ad a \rangle = ( e^{\beta w } -1 ) \left[ \frac {\mu}{
 ( e^{\beta w } -q^{-1})^2 } +   \frac {1-\mu}{ ( e^{\beta w } -q  )^2 } \right].
\ee
As seen in (\ref{eq13}) and in the formula (\ref{swh}),
this expression of one-particle distribution (also called mean occupation number) separates into two terms by putting
$\mu=0,$ or $\mu=1,$ respectively, which are nothing but ordinary Tamm-Dancoff formulae (though with  $q\to 1/q$ in case of $\mu=1$). On the other hand, the one-particle distribution  formula for the usual Tamm-Dancoff model readily follows, at $p\to q,$  of  \cite{Daoud:1995eu}.
%
 Besides, when $ q \rightarrow 1,$ the  equation (\ref{swh}) reduces
to the classical result of nondeformed case, i.e.,
\be
\langle \ad a \rangle = \frac {1}{e^{ w /kT } -1 }.
\ee
It also appears that the 
mean occupation number for deformed 
photons
 obeying the GTD-algebra
has a discontinuity at $ x = \ln \qi $ for $0<q<1$, where $x = \beta w $. Figure 1 shows the 
discontinuity of $\langle \ad a \rangle $. The discontinuity disappears if we restrict 
the range of $x$ to $ x_{min }= \ln \qi < x < \infty $. Figure 2 shows the 
distribution without discontinuity.

Now let us discuss the black body radiation for deformed photons that obey the 
algebra {\color{blue} (\ref{eq13})}. The mean energy for STD photons is the energy of 
single GTD photon multiplied by the mean occupation number as follows:
\be
\langle E \rangle = w \bar{n} = w ( e^{\beta w } -1 ) \left[ \frac {\mu}{
 ( e^{\beta w } -q^{-1} )^2 } +   \frac {1-\mu}{ ( e^{\beta w } -q )^2 } \right],
\ee
where $ w = \h \nu $ and $ \nu $ is a frequency of STD photon. The total 
energy per unit volume for GTD photons in the cavity is obtained by
\be
\label{total:energy}
U(T) = \frac{ 8 \pi \h }{c^3 } \int_0^{\infty} d \nu\,  \nu^3 ( e^{\beta \h\nu } -1 ) 
\left[ \frac {\mu}{ ( e^{\beta \h \nu } -q^{-1} )^2 } +   \frac {1-\mu}{ ( e^{\beta \h \nu } -q )^2 } \right].
\ee
If we set $ x = \beta \h \nu $, we have
\be
U(T) = \frac{ 8 \pi \h }{c^3 } \lb \frac{ kT} {\h} \rb^4 \int_0^{\infty} d x \, 
x^3  ( e^{x } -1 ) \left[ \frac {\mu}{ ( e^{x } -q^{-1} )^2 } +   \frac {1-\mu}{ ( e^{x } -q )^2 } \right] = a_q T^4,
\ee
which is the deformed Stefan-Bolzmann law; indeed, the proportional 
coefficient is $q-$deformed. Now let us calculate the integral part;
\be
\label{average:s}
J(q) = \int_0^{\infty} d x\, x^3  ( e^{x } -1 )
 \left[ \frac {\mu}{ ( e^{x } -q^{-1} )^2 } +   \frac {1-\mu}{ ( e^{x } -q  )^2 } \right]
= 12 \sum_{n=0}^{\infty} \{ n \}_\mu  \lb \frac{\mu}{(n+1)^4 } - \frac{1-\mu}{(n+2)^4 } \rb,
\ee
where we can easily find that  $ J(q)\rightarrow 6 \,\zeta (4) = \frac { \pi^4 }{15} $
 when $ q $ goes to  unity and $ \mu = 1/2 $ and $\zeta(.)$ is the   Riemann zeta 
function. The infinite series given in (\ref{average:s}) diverges. It results from the 
discontinuity of the mean occupation number of GTD photons. To resolve this 
problem, we should restrict the range of 
$x$ to $ x_{min } < x $. In this case, the integral part is changed into
\be
J(q) = \int_{\ln \qi }^{\infty} d x \, x^3 ( e^{x } -1 ) 
\left[ \frac {\mu }{ ( e^{x } -q^{-1} )^2 } +   \frac {1-\mu}{ ( e^{x } -q)^2 } \right].
\ee
From the (\ref{total:energy}), the average energy per mode $ I ( \nu ) $ is given by
\be
I ( \nu) = \frac{ 8 \pi \h \nu^3  }{c^3 }   ( e^{\frac{ \h \nu}{ kT}} -1 ) 
\left[ \frac {\mu }{ ( e^{\frac{ \h \nu}{ kT} } -q^{-1} )^2 } +  
 \frac {1-\mu }{ ( e^{\frac{ \h \nu}{ kT} } -q )^2 } \right].
\ee
 Figures 1- 3 show the plots of $I(\nu)$ with  $ x> x_{min} $ for 
$ q =0.78, \mu=0.1, 0.5, 0.9$ (continuous line) and for $ q=1$ (dashed line). 
One can observe that, for the considered deformation parameter values,   the deformed   
average energy per mode $ I( \nu ) $ for $q=0.78$  increases between $\nu=0$ and $\nu=4$ while remaining under  the values of the non-deformed case ($q=1$) as  $\mu $ increases, and maintains the same decreasing trend as the non-deformed case for $\nu\ge 4.$
\newpage
\begin{figure}
\includegraphics[width=7.4cm]{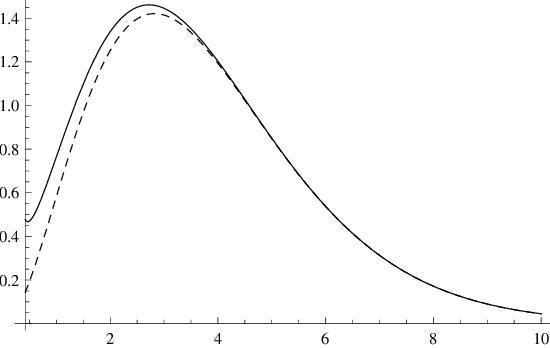}\\
{\bf Figure 1.} {Plot of $I(\nu)$ with $ x> x_{min} $ for $ q =0.78, \mu=0.1$ (continuous line) and for $ q=1$ (dashed line)}\\
\includegraphics[width=7.4cm]{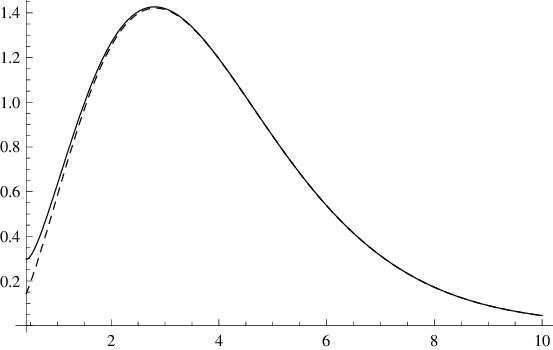}\\
{\bf Figure 2.} {Plot of $I(\nu)$ with $ x> x_{min} $ for $ q =0.78, \mu=0.5$ (continuous line) and for $ q=1$ (dashed line)}\\
\includegraphics[width=7.4cm]{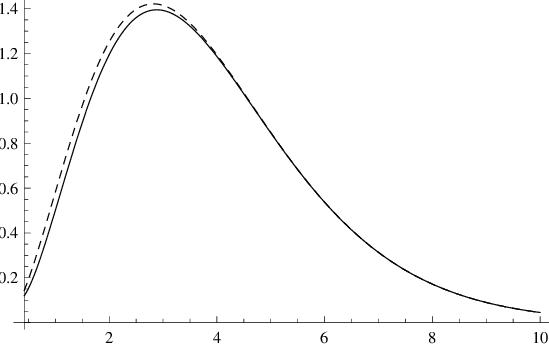}\\
{\bf Figure 3.} {Plot of $I(\nu)$ with $ x> x_{min} $ for $ q =0.78, \mu=0.9$ (continuous line) and for $ q=1$ (dashed line).}
\end{figure}

\section{Conclusion}
In this work, we have proposed a   full characterization of a generalized
TD-oscillator algebra and  investigated its main mathematical and physical properties.
Specifically, we have studied its various  representations
and found the 
condition satisfied by  the deformed $q-$number to define the algebra structure function. Particular Fock spaces involving   finite and infinite dimensions have been examined.  A
 deformed calculus has been  performed as well as  a coordinate realization for this 
algebra. A relevant example of the generalized $q-$deformed 
TD oscillator algebra
has been exhibited. Associated
 coherent states have been constructed with required mathematical conditions. 
Besides,   some thermodynamics aspects have been computed.

Finally, let us mention that, although  the main part of this work dealt with only two parameters
$q$ and $\mu,$ the investigation of  the more general case with a number of parameters greater than two
can be performed in a similar way as done in \cite{Hounk&Elvis}. For instance, the multi-parameter deformed algebra 
 (\ref{deformed:al}) and  its deformed number (\ref{number:d}) lead to the following actions for the operators $N,\,a$ and $\ad$ on the Fock space, for $ n=0, 1, 2, \cdots$: 
\bea
N|n\rangle &=& n |n\rangle,\\
a|n\rangle &=& \sqrt{  \{n\}     }\,|n-1\rangle,\\
\ad |n\rangle& =& \sqrt{\{n+1\}  }\,|n+1\rangle,
\eea
where
\be
\{n\}= n \lb  \mu q^{ \alpha n + \beta } + \eta 
q^{ \gamma  n + \delta }  \rb.
\ee
In this case, the position  and momentum operators $Q$ and $P,$ 
\be
Q:= \left(1/{2 \, m}\,\omega\right)^{1/2}(a^\dag+a)\quad 
\mbox{ and } \quad P:=i \left(m \, \omega/{2}\right)^{1/2}(a^\dag-a),
\ee
allow us to define  the  GTD oscillator Hamiltonian operator $H$ and its eigenvalue as:
\bea
H:
=\frac{1}{2m}P^2+\frac{1}{2}m\omega^2Q^2&=&\frac{ \omega}{2}(a^\dagger a+aa^\dagger)\cr
&=&\frac{  \omega}{2}\Big(\{N\}+ \{N+1\}\Big)
\eea
and 
\be
 E(n)=  \frac{  \omega}{2}\Big(\{n\}+ \{n+1\}\Big),
\ee
respectively.
All other results obtained in this work can be formally extended to the multiparameter deformation case, except for the resolution of the moment problem which can be a difficult task. A thorough analysis of all these questions will be in the core of the forthcoming paper.


%
\def\JMP #1 #2 #3 {J. Math. Phys. {\bf#1},\ #2 (#3).}
\def\JP #1 #2 #3 {J. Phys. A {\bf#1},\ #2 (#3).}
\def\JPMT #1 #2 #3 {J. Phys. A: Math. Theor. {\bf#1},  #2 (#3).}
\def\PLA #1 #2 #3 {Phys. Lett.  A {\bf#1},\ #2 (#3).}
\def\PLB #1 #2 #3 {Phys. Lett.  B {\bf#1},\ #2 (#3).}
\def\JMA #1 #2 #3 {Mod. Phys. Lett. A {\bf#1},\ #2 (#3).}
\def\JMB #1 #2 #3 {Mod. Phys. Lett. B {\bf#1},\ #2 (#3).}
\def\JM #1 #2 #3 {Mess. Math. {\bf#1},\ #2 (#3).}
\def\LMP #1 #2 #3 {Lett. Math. Phys. {\bf#1},\ #2 (#3).}

\end{document}